\documentclass[prl,reprint,
preprintnumbers]{revtex4-2}
\pdfoutput=1
\usepackage{hyperref}
\usepackage[T1]{fontenc}

\usepackage{mathrsfs}

\usepackage[inline]{enumitem}

\usepackage{amsmath,amssymb,amsfonts,amsxtra,mathrsfs,graphics,graphicx,amsthm,epsfig,bm,longtable,float,color,tikz,mathtools,xfrac,footnote}
\restylefloat{table}
\usepackage{dsfont}
\usepackage{lipsum}
\usepackage{textgreek}
\usetikzlibrary{decorations.pathmorphing}
\usetikzlibrary{decorations.markings}
\usetikzlibrary{quotes,arrows.meta}
\usetikzlibrary{arrows,decorations.markings,calc,fadings,decorations.pathreplacing,patterns,decorations.pathmorphing,positioning}
\usepackage{tikz-cd}

\newcommand{\ii}{\mathrm{i}}
\newcommand{\?}{\;\!}

\newcommand{\nn}{\nonumber}

\newcommand{\be}{\begin{equation}} \newcommand{\ee}{\end{equation}}
\newcommand{\bea}{\begin{equation} \begin{aligned}} \newcommand{\eea}{\end{aligned} \end{equation}}

\def\U{\mathrm{U}}

\def\SU{\mathrm{SU}}

\newcommand{\rd}{\mathrm{d}}

\usepackage{relsize}
\newcommand{\Vol}{\mathlarger{\mathrm{Vol}}}

\DeclareMathOperator{\Tr}{Tr}

\DeclareMathOperator{\Li}{Li}

\newcommand{\cC}{\mathcal{C}}

\newcommand{\cF}{\mathcal{F}}
\newcommand{\cG}{\mathcal{G}}

\newcommand{\cN}{\mathcal{N}}
\newcommand{\cO}{\mathcal{O}}

\newcommand{\cR}{\mathcal{R}}

\newcommand{\bZ}{\mathbb{Z}}

\usepackage[bbgreekl]{mathbbol}
\DeclareMathSymbol\bbDelta  \mathord{bbold}{"01}

\makeatletter
\makeatletter

\usepackage{orcidlink}

\begin{document}

\title{Large-$N$ Free Energy of Chiral $\mathcal{N}=2$ Chern-Simons-Matter Theories}

\date{\today}

\author{Seyed Morteza Hosseini\orcidlink{0000-0001-8205-400X}}
\email{morteza.hosseini@qmul.ac.uk}
\affiliation{Centre for Theoretical Physics, Department of Physics and Astronomy, Queen Mary University of London, London E1 4NS, UK}

\begin{abstract}
We present the first successful large-$N$ computation of the $S^3$ free energy in chiral $\mathcal{N}=2$ Chern-Simons-matter theories, long believed to evade the universal M2-brane scaling $F_{S^3}\!\sim\!N^{3/2}$. 
Using a stable numerical continuation method that directly solves the saddle-point equations, we obtain convergent large-$N$ solutions for benchmark chiral quivers, including the holographic dual to AdS$_4\times Q^{1,1,1}/\mathbb{Z}_k$ and $D_3$ models. 
The resulting free energies precisely match the holographic predictions, resolving a decade-old puzzle in AdS$_4$/CFT$_3$ and establishing a practical numerical framework for precision holography in strongly coupled gauge theories.
\end{abstract}



\maketitle

\emph{Introduction---}%
The large-$N$ behavior of partition functions of three-dimensional superconformal field theories (SCFTs) describing the dynamics of $N$ coincident M2-branes provides one of the most stringent precision tests of the AdS$_4$/CFT$_3$ correspondence. 
For $\mathcal{N}=2$ Chern-Simons-matter gauge theories, supersymmetric localization renders the $S^3$ partition function exactly computable as a finite-dimensional matrix integral \cite{Kapustin:2009kz,Jafferis:2010un,Hama:2010av}. 
At large $N$, the free energy $F \equiv -\log Z$ is expected to exhibit the universal M2-brane scaling
\begin{equation}
\label{AdS:CFT}
F = N^{3/2} \sqrt{\frac{2\pi^6}{27 \? \Vol(Y_7)}},
\end{equation}
in agreement with the holographic prediction from M-theory on AdS$_4\times Y_7$, where $Y_7$ is a seven-dimensional Sasaki-Einstein manifold and AdS$_4$ refers to four-dimensional anti-de Sitter space \cite{Drukker:2010nc,Herzog:2010hf}.

Many such theories arise as Chern-Simons deformations of four-dimensional quiver gauge theories describing D3-branes at Calabi-Yau threefold singularities \cite{Aharony:2008ug,Hanany:2008cd,Hanany:2008fj,Martelli:2008si,Franco:2008um,Davey:2009sr,Hanany:2009vx,Franco:2009sp}, or through the addition of fundamental flavors \cite{Gaiotto:2009tk,Jafferis:2009th,Benini:2009qs}.
These constructions lead to two broad classes of Chern-Simons-matter models:
\begin{enumerate*}[label=(\roman*)]
 \item \emph{nonchiral} (vectorlike) quivers, in which every bifundamental chiral multiplet is accompanied by its conjugate, and
 \item \emph{chiral} quivers, whose matter content lacks charge-conjugation symmetry.
\end{enumerate*}

For nonchiral theories, the large-$N$ matrix model is well understood: long-range forces among eigenvalues cancel, producing a smooth saddle and a free energy in quantitative agreement with the supergravity result \cite{Drukker:2010nc,Herzog:2010hf,Jafferis:2011zi,Martelli:2011qj,Cheon:2011vi,Gulotta:2011si,Gulotta:2011vp}.
In contrast, applying the same techniques to chiral quivers leads to persistent difficulties.
Long-range forces fail to cancel, the eigenvalue distribution becomes irregular, and the resulting free energy scales as $N^2$ rather than $N^{3/2}$ \cite{Herzog:2010hf}.
Although these theories are well defined and pass nontrivial consistency checks at finite $N$ \cite{Benini:2011cma}, no consistent large-$N$ solution reproducing the expected scaling had been found.
This long-standing mismatch cast doubt on whether chiral quivers admit a well-defined large-$N$ saddle compatible with holography, or whether additional ingredients are required to ``repair'' the matrix model.
One proposed remedy is to symmetrize the matrix integrand \cite{Amariti:2011jp,Amariti:2011uw,Gang:2011jj}, but this prescription does not follow from localization and does not correspond to a genuine saddle of the original integral.
The status of the large-$N$ limit in chiral quivers therefore remains unresolved.

In this Letter, we overcome this impasse by introducing a robust numerical framework that directly solves the finite-$N$ saddle-point equations of the $S^3$ matrix model.
Earlier approaches interpreted these equations as the equilibrium conditions of interacting particles in the complex plane and evolved them via gradient-flow (heat-equation) relaxation \cite{Herzog:2010hf}.
While effective for vectorlike theories, this method fails for chiral quivers due to stiff, nonconvergent dynamics arising from complex interaction forces.
Our approach combines Anderson acceleration with an adiabatic continuation in $N$: starting from a convergent low-$N$ solution, we increase $N$ step by step, solving at each stage with a Levenberg-Marquardt or Powell-hybrid update and refining via a damped Newton iteration with line search.
The initial guess for $N + 1$ is generated by monotone interpolation of eigenvalues and a $\sqrt{(N + 1)/N}$ rescaling of their real parts, preserving the anticipated M2-brane scaling.
This continuation algorithm converges reliably across the tested $(N , k)$ range and yields precise free energies even in regimes where relaxation methods fail.

We apply this framework to two benchmark examples: the quiver dual to AdS$_4\times Q^{1,1,1}/\mathbb{Z}_k$ and the four-node $D_3$ quiver of \cite{Franco:2008um}.
In both cases, we find well-defined large-$N$ saddles whose free energies scale as $N^{3/2}$ and match the holographic prediction \eqref{AdS:CFT}.
These results provide the \emph{first} quantitative confirmation of the AdS$_4$/CFT$_3$ correspondence for chiral quivers, resolving a decade-old puzzle and establishing a practical numerical framework for precision holography for strongly coupled SCFTs beyond the vectorlike realm.

\emph{The three-sphere matrix model---}%
We consider a three-dimensional $\cN=2$ gauge theory with gauge group $\cG=\prod_{a=1}^{|\cG|}\cG_a$, Chern-Simons levels $k_a$ for each factor, and chiral multiplets in representations $\bigoplus_{I}\cR_I$ of $\cG$.
By supersymmetric localization, the $S^3$ partition function, viewed as a function of trial R-charges $\Delta$, reduces to a matrix integral over the Cartan of $\cG$ \cite{Jafferis:2010un,Hama:2010av}
\bea
 \label{Z:S3}
 Z ( \Delta ) & = \int_{-\infty}^{\infty}
 \Biggl( \prod_{a = 1}^{|\cG|} \prod_{i = 1}^{\text{rk} ( \cG_a )} \rd  \lambda_{a,i} \Biggr)
 \? \exp \left( - \cF ( \lambda_{a, i}, \Delta ) \right) \, .
\eea
The free energy functional is
\bea
 \label{F:S3}
 \cF ( \lambda_{a,i} , \Delta ) & = \log |W| + \text{rk} ( \cG ) \log( 2 \pi)
 - \ii \sum_{a=1}^{|\cG|} \! \sum_{i=1}^{\text{rk} ( \cG_a )} \! \frac{k_a}{4 \pi} \lambda_{a,i}^2 \\
 & + \sum_{a=1}^{|\cG|} \! \sum_{i=1}^{\text{rk} ( \cG_a )} \! \Delta_{m}^{(a)} \lambda_{a, i}
 - 2 \sum_{ \alpha \in \bbDelta_{+} } 
 \log\!\left( 2 \sinh \frac{ \alpha ( \lambda ) }{2} \right) \\
 & - \sum_{I} \! \sum_{\rho_I \in \cR_I}
 \ell \! \left( 1 - \Delta_{I} + \ii \frac{ \rho_I ( \lambda )}{ 2 \pi } \right) .
\eea
Here $\alpha$ runs over the positive roots of $\cG$, $\rho_I$ are the weights in $\cR_I$, and $|W|$ is the order of the Weyl group.
The parameters $\Delta_m^{(a)}$ arise from $\U(1)$ topological symmetries associated with the conserved currents $\star\!\Tr F_a$, which couple linearly to the eigenvalues $\lambda_{a,i}$; in this Letter we set $\Delta_m^{(a)}=0$.  
The function
\be
 \ell (z ) = \frac{\ii}{2} \left( \frac{1}{\pi} \Li_2 \left( e^{2 \pi \ii z} \right) + \pi z^2 \right) + z \Li_1\left( e^{2 \pi \ii z} \right) -\frac{\ii \pi}{12} \nn \, ,
\ee
satisfies
$
 \frac{\rd \ell ( z )}{\rd z} = - \pi z \cot ( \pi z ) \, .
$
Eigenvalues $\lambda_{a,i}$ are integrated along the real axis, and we use the principal branches of the polylogarithms in $\ell(z)$ with $0<\Delta_I<1$.
In the large-$N$ limit, the integral \eqref{Z:S3} is dominated by a saddle point of the effective action in the exponent.
Writing $Z ( \Delta )=e^{- F ( \Delta )}$, the free energy $F ( \Delta )$ is obtained by extremizing \eqref{F:S3} with respect to all eigenvalues $\lambda_{a,i}$,
\bea
 \label{eq:saddle-point}
 \frac{\partial \cF(\lambda_{a,i},\Delta)}{\partial\lambda_{a,i}}=0 \, .
\eea
These coupled saddle-point equations determine the large-$N$ eigenvalue distributions for each gauge group and are solved numerically in the examples below.

\emph{Holographic dual to AdS$_4 \times Q^{1,1,1}/\bZ_k$---}%
As a first example, we consider the chiral Chern-Simons-matter theory describing the worldvolume dynamics of $N$ M2-branes at the tip of the cone over $Q^{1,1,1}/\bZ_k$, where $Q^{1,1,1}$ is the coset space $\SU(2)^3/\U(1)^2$.
At large $N$, this theory is conjectured to be holographically dual to M-theory on AdS$_4 \times Q^{1,1,1}/\bZ_k$ \cite{Franco:2008um,Franco:2009sp}.
Using $\mathcal{N}=2$ notation, the matter content is described by the quiver diagram \cite{Franco:2008um}:
\bea
\label{Q111:quiver}
\begin{tikzpicture}[
  scale=0.75,
  transform shape,
  gauge/.style={circle,fill=none,draw=black,minimum size=10mm,inner sep=0pt,
                text=black,font=\normalsize},
  edge/.style={-Latex,thick},
  edgelabel/.style={midway,sloped,above,text=black,font=\normalsize}
]
  \node[gauge] (L) at (-2.2,0) {$N_k$};
  \node[gauge] (R) at ( 2.2,0) {$N_{-k}$};
  \node[gauge] (T) at ( 0,1.6) {$N_0$};
  \node[gauge] (B) at ( 0,-1.6) {$N_0$};

  \draw[edge, ->] (L) -- node[edgelabel]{$A_1$} (T);
  \draw[edge, ->] (R) -- node[edgelabel]{$A_2$} (T);
  \draw[edge, ->] (B) -- node[edgelabel,below]{$C_1$} (L);
  \draw[edge, ->] (B) -- node[edgelabel,below]{$C_2$} (R);
  \draw[edge, ->>] (T) -- node[right,text=black,font=\normalsize]{$B_1, B_2$} (B);
\end{tikzpicture}
\eea
Each node denotes a unitary gauge group $\U(N)_k$, where the subscript indicates the Chern-Simons level \footnote{In \cite{Franco:2008um}, the Abelian case $N = 1$ with Chern-Simons levels $\vec{k}=(1,-1,0,0)$ was shown, via toric geometry, to yield the moduli space $\cC(Q^{1,1,1})$. The alternative assignment $\vec{k}=(1,1,-1,-1)$ gives the same moduli space; here we focus on $\vec{k}=(k,-k,0,0)$.}. 
A bifundamental field $\Phi_{a \to b}$ transforms in the fundamental representation of node $a$ and the antifundamental representation of node $b$.
The arrows represent chiral multiplets $A_r$, $B_r$, and $C_r$ in bifundamental representations of adjacent nodes, following this convention.
Up to normalization, the superpotential reads
\be
 W = \Tr \left( C_1 A_1 B_1 C_2 A_2 B_2 - C_1 A_1 B_2 C_2 A_2 B_1 \right) \nn \, .
\ee
For $k=1$, the global symmetry is $\SU(2)^3 \times \U(1)_R \times \U(1)_B^2$. 
The pairs $(A_1,A_2)$, $(B_1,B_2)$, and $(C_1,C_2)$ transform as doublets under the three $\SU(2)$ factors, while the two $\U(1)_B$ are baryonic symmetries. 
Marginality of the superpotential requires
$
 \sum_{r = 1}^{2} \left( \Delta_{A_r} + \Delta_{B_r} + \Delta_{C_r} \right) = 2.
$
Exploiting the symmetries of the quiver, we set $\Delta_{A_r} = \Delta_{B_r} = \Delta_{C_r} \equiv \Delta_r$, 
which leads to
$
 \sum_{r = 1}^{2} \Delta_r = 2/3.
$

We compute the Sasakian volume of $Q^{1,1,1}/\bZ_k$ as a function of the trial R-charges $\Delta_r$ using the toric construction of \cite{Martelli:2005tp,Martelli:2006yb}. This yields
$
 \Vol(Q^{1,1,1}/\bZ_k) = \pi^4/\left[72 k \Delta_1 \Delta_2 \left( 2 \Delta_1 + \Delta_2 \right) \left( \Delta_1 + 2 \Delta_2 \right)\right]
$,
which is minimized at $\Delta_r = \tfrac{1}{3}$, giving
$
 \Vol(Q^{1,1,1}/\bZ_k)\big|_{\text{min}} = \pi^4/(8k) .
$
Substituting this into the AdS$_4$/CFT$_3$ relation for the $S^3$ free energy \eqref{AdS:CFT}, yields the holographic prediction
\be
 \label{Q111:c32:an}
 F_{Q^{1,1,1}/\bZ_k} = \frac{4\pi}{3\sqrt{3}} \? k^{1/2} N^{3/2} \approx 2.418 \? k^{1/2} N^{3/2} \, .
\ee
This value will serve as the benchmark for comparison with the numerical results discussed below.

\emph{The $D_3$ theory---}%
As a second example, we consider the $D_3$ Chern-Simons-matter gauge theory~\cite{Franco:2008um}.
The quiver diagram is shown below
\bea
\label{D3:quiver}
\begin{tikzpicture}[
  scale=0.75,
  transform shape,      
  gauge/.style={circle,fill=none,draw=black,minimum size=10mm,inner sep=0pt,
                text=black,font=\normalsize},
  edge/.style={-Latex,thick},
  edgelabel/.style={midway,sloped,above,text=black,font=\normalsize}
 ]
  \node[gauge] (L) at (-2.2,0) {$N_k$};
  \node[gauge] (R) at ( 2.2,0) {$N_{k}$};
  \node[gauge] (T) at ( 0,1.6) {$N_{-k}$};
  \node[gauge] (B) at ( 0,-1.6) {$N_{-k}$};

  \draw[edge, ->] (L) -- node[edgelabel]{$A_1$} (T);
  \draw[edge, <-] (R) -- node[edgelabel]{$A_2$} (T);
  \draw[edge, ->] (B) -- node[edgelabel,below]{$C_1$} (L);
  \draw[edge, <-] (B) -- node[edgelabel,below]{$C_2$} (R);
  \draw[edge, <->] (T) -- node[right, text=black, font=\normalsize]{$B_1, B_2$} (B);

\end{tikzpicture}
\eea
Each node denotes a unitary gauge group $\U(N)_k$ with the indicated Chern-Simons level \footnote{Similar to the $Q^{1,1,1}/\bZ_k$ case, two distinct Chern-Simons level assignments, $\vec{k} = (k, k, -k, -k)$ and $\vec{k} = (k, -k, 0, 0)$, yield the same moduli space. Here we focus on $\vec{k} = (k, k, -k, -k)$.}, and the arrows represent chiral multiplets in bifundamental representations following the same convention for $\Phi_{a \to b}$ defined above.
Up to a normalization, the superpotential reads
\be
 W = \Tr \left( C_1 A_1 B_1 B_2 A_2 C_2 - B_1 C_1 A_1 A_2 C_2 B_2 \right) \nn \, .
\ee
For $k=1$, the global symmetry is $\U(1)^3 \times \U(1)_R \times \U(1)_B^2$, where the $\U(1)_B$ factors correspond to baryonic symmetries. 
Marginality of the superpotential imposes the constraint
$
 \sum_{r = 1}^{2} \left( \Delta_{A_r} + \Delta_{B_r} + \Delta_{C_r} \right) = 2.
$

Using the toric construction of \cite{Martelli:2005tp,Martelli:2006yb}, the volume of the dual Sasakian manifold takes the form
$
\Vol(D_3)
= \pi^4 / \bigl[ 24k \prod_{l=1}^{5} X_l \bigr],
$
where
$
X_1 \equiv \Delta_{C_1}+\Delta_{C_2} ,
$
$
X_2 \equiv \Delta_{A_2}+\Delta_{B_1} ,
$
$
X_3 \equiv \Delta_{A_1}+\Delta_{B_2} ,
$
$
X_4 \equiv \Delta_{A_1}+\Delta_{A_2}+\Delta_{C_1} ,
$
$
X_5 \equiv \Delta_{B_1}+\Delta_{B_2}+\Delta_{C_2} .
$
Minimization with respect to the trial R-charges yields
$
X_1=X_2=X_3=2/3
$
and
$
X_4=X_5=1 ,
$
which gives
$
\Vol(D_3)\big|_{\text{min}} = 9\pi^4 / (64k) .
$
Substituting this into \eqref{AdS:CFT} gives the holographic prediction for the $S^3$ free energy,
\be
 \label{D3:c32:an:can}
 F_{D_3} = \frac{8 \pi}{9} \sqrt{\frac{2}{3}} k^{1/2} N^{3/2} \approx 2.280 \? k^{1/2} N^{3/2} \, ,
\ee
which serves as a benchmark for comparison with the numerical results presented below.
As an additional test away from the canonical R-charge assignment, we consider
$\Delta_{A_1}=0.12$,
$\Delta_{A_2}=0.18$,
$\Delta_{C_2}=0.27$,
$\Delta_{C_1}=0.33$,
$\Delta_{B_1}=0.41$, and
$\Delta_{B_2}=0.69$,
which satisfies the marginality constraint but does not extremize the free energy.
Evaluating $\Vol(D_3)$ for this choice and substituting into \eqref{AdS:CFT} yields
\be
\label{D3:c32:an:non-can}
F_{D_3}^{(\Delta)} \approx 2.084\, k^{1/2} N^{3/2} \, ,
\ee
which we use as a non-canonical benchmark for comparison with the numerical results.

\emph{Numerical method---}%
The saddle-point equations \eqref{eq:saddle-point} define a coupled nonlinear system for the complex eigenvalues $\lambda_{a,i}$.
We rewrite them as a residual map
\be
\begin{aligned}
 \label{eq:residual}
 R_{a,i}(\lambda,\Delta)
 & \equiv - \frac{\partial \cF(\lambda,\Delta)}{\partial \lambda_{a,i}}
 = \frac12 \sum_{\alpha \in \text{Adj}_a} \alpha^{(a,i)} \coth\!\frac{\alpha(\lambda)}{2} \\
 & - \frac{\ii}{2} \sum_{I}\sum_{\rho_I\in\cR_I} \rho_I^{(a,i)}\, z_I \cot(\pi z_I)
 + \frac{\ii k_a}{2\pi} \lambda_{a,i} \, ,
\end{aligned}
\ee
where
$
 z_I \equiv 1 - \Delta_I + \frac{\ii}{2\pi}\rho_I(\lambda) .
$
Here $\alpha^{(a,i)}$ and $\rho_I^{(a,i)}$ denote, respectively, the components of the root $\alpha$ and the weight $\rho_I$ along the Cartan direction associated with $\lambda_{a,i}$.
The large-$N$ saddle corresponds to $R_{a,i}(\lambda,\Delta)=0$ for all $a,i$.

We solve \eqref{eq:residual} directly at finite $N$ using a robust nonlinear scheme that combines Anderson acceleration with a Levenberg-Marquardt or Powell-hybrid step and a damped Newton refinement with a backtracking line search \footnote{All solvers are standard routines from \textsc{SciPy}'s \texttt{optimize.root} module: Anderson acceleration (\texttt{anderson}), Levenberg-Marquardt (\texttt{lm}), and Powell-hybrid (\texttt{hybr}). The damped Newton refinement with backtracking line search is a custom implementation of a standard Jacobian-based method used to improve convergence.}.
At each fixed $N$, Anderson acceleration is applied to the fixed-point map
$\lambda \mapsto \lambda - R ( \lambda , \Delta )$.
This iterative scheme proves markedly more stable than gradient-flow or relaxation approaches, which fail for chiral quivers due to oscillatory complex forces and the absence of a dissipative direction.
Anderson acceleration, by contrast, converges reliably to a stationary complex saddle over a broad range of initial conditions.
Starting from a low-$N$ seed, it identifies a smooth branch suitable for continuation to larger~$N$.
Convergence is declared once the residual norm $\|R\|_2 \equiv \sqrt{ \sum_i R_i^2 }$ falls below $10^{-12}$, or when the relative update becomes numerically stationary,
$\| \Delta \lambda \|_2 \le10^{-14}( 1 + \| \lambda \|_2)$.

To access the large-$N$ regime, we employ an adiabatic continuation in $N$.
Starting from a convergent low-$N$ configuration, we increase $N$ step by step.
The seed for $N + 1$ is generated by monotone interpolation of the ordered eigenvalues $\{\lambda_{a,i}\}$,
\bea
 \mathrm{Re}\lambda_{a,i}^{\,\text{seed}}
 & = \overline{\mathrm{Re}\lambda}_a
 + \sqrt{\tfrac{N{+}1}{N}}\bigl(\mathrm{Re}\lambda_{a,i}^{\,\text{interp}}-\overline{\mathrm{Re}\lambda}_a\bigr) \nn \, ,
 \\
 \mathrm{Im}\lambda_{a,i}^{\,\text{seed}} & = \mathrm{Im}\lambda_{a,i}^{\,\text{interp}} \, ,
\eea
where $\overline{\mathrm{Re}\lambda}_a$ is the mean real part for node $a$
and $\mathrm{Re}\lambda_{a,i}^{\,\text{interp}}$ denotes the monotone interpolation to $N +1$ points.
The $\sqrt{(N{+}1)/N}$ factor preserves the expected $\sqrt{N}$ widening of
eigenvalue support implied by $F\!\sim\!N^{3/2}$,
while the imaginary parts are continued unchanged.
This continuation, in conjunction with Anderson acceleration, provides a reliable route to smooth large-$N$ saddles within the numerically accessible range.

The on-shell free energy
$F ( \Delta ) \equiv \cF(\lambda^\star,\Delta)$
is evaluated including the measure term $\log|W|+\text{rk}(\cG)\log(2\pi)$.
We report $\mathrm{Re}F$ as the physical value and monitor $\mathrm{Im}F$ as a diagnostic of numerical accuracy.
To extract the leading scaling coefficient, we perform a least-squares fit of $\mathrm{Re} F(N)$
to the basis
$\{N^{3/2},\,N\log N,\,N,\,N^{1/2},\,\log N,\,1\}$;
the coefficient of $N^{3/2}$ defines the numerical $c_{3/2}$.

\begin{figure}[htp!]
  \centering
  \includegraphics[width=\columnwidth]{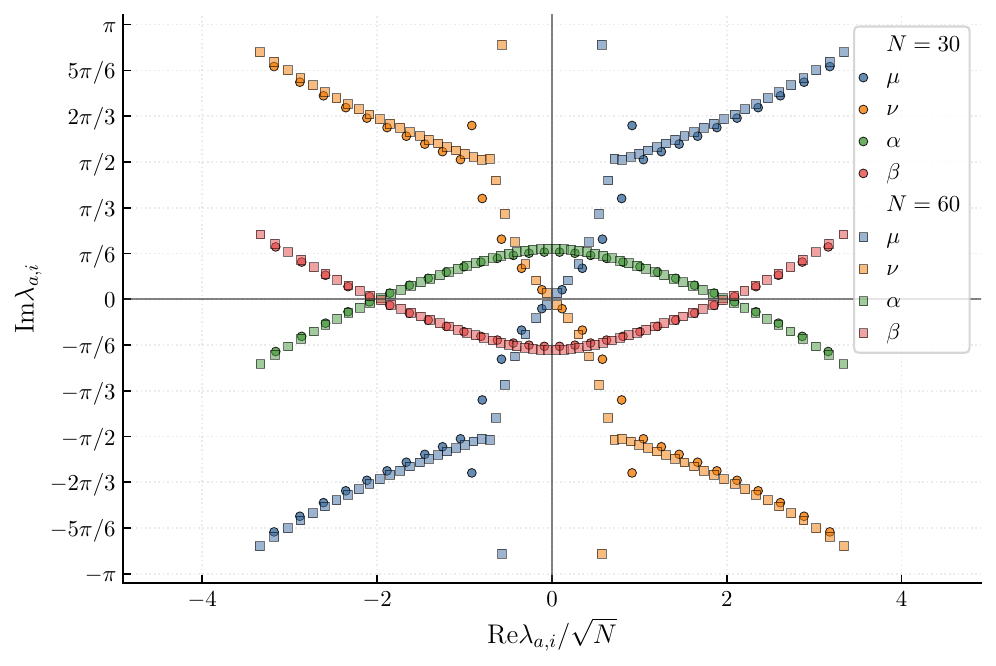}
    \caption{
        Complex eigenvalue distributions for the $Q^{1,1,1}/\mathbb{Z}_k$ quiver at $N=30$ and $N=60$.
        Colors label the gauge nodes:
        $\U(N)_k$ ($\mu_i$), $\U(N)_{-k}$ ($\nu_i$),
        $\U(N)_0^{(T)}$ ($\alpha_i$), and $\U(N)_0^{(B)}$ ($\beta_i$).
        Each node forms a smooth curve; real parts scale as $\sqrt{N}$, while imaginary parts remain $\cO(1)$ and cluster into node-dependent bands.
        The distributions satisfy $\nu_i = \overline{\mu_i}$ and $\beta_i = \overline{\alpha_i}$, indicating complex conjugation symmetry across the quiver.
    A few points at $N=60$ lie slightly off the smooth curves; these do not affect any of the quantities reported.
    }
  \label{fig:Q111_eigs_complex}
\end{figure}

\emph{Results and discussion---}%
We applied the adiabatic continuation with Anderson acceleration to the chiral $Q^{1,1,1}/\mathbb{Z}_k$ and $D_3$ quiver theories; all numerical data use $k=1$.
Figure~\ref{fig:Q111_eigs_complex} shows the complex eigenvalue distributions for
$Q^{1,1,1}/\mathbb{Z}_k$ at $N=30$ and $N=60$.
The four branches correspond to the eigenvalue sets $\{\mu_i\}$, $\{\nu_i\}$, $\{\alpha_i\}$, and $\{\beta_i\}$, associated, respectively, with the $\U(N)_k$, $\U(N)_{-k}$, and the two $\U(N)_0$ nodes of the quiver~\eqref{Q111:quiver}, denoted $\U(N)_0^{(T)}$ (top) and $\U(N)_0^{(B)}$ (bottom).
Each set forms a smooth curve in the complex plane.
As $N$ increases, the real parts broaden as $\propto \sqrt{N}$
while the imaginary parts remain $\mathcal{O}(1)$ and arrange into node-dependent bands.
The distributions exhibit complex conjugation symmetry:
$\nu_i = \overline{\mu_i}$ and $\beta_i = \overline{\alpha_i}$,
reflecting the quiver's structural symmetry.
The smooth continuation of these branches confirms that the numerical solver
tracks the same physical saddle as $N$ increases.

\begin{figure}[htp!]
  \centering
  \includegraphics[width=\columnwidth]{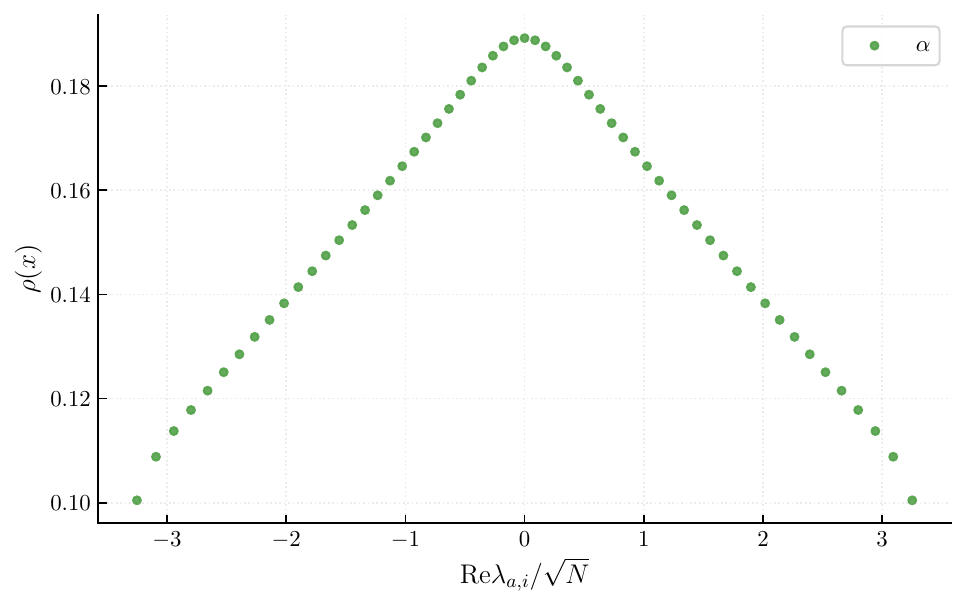}\\[2pt]
  \includegraphics[width=\columnwidth]{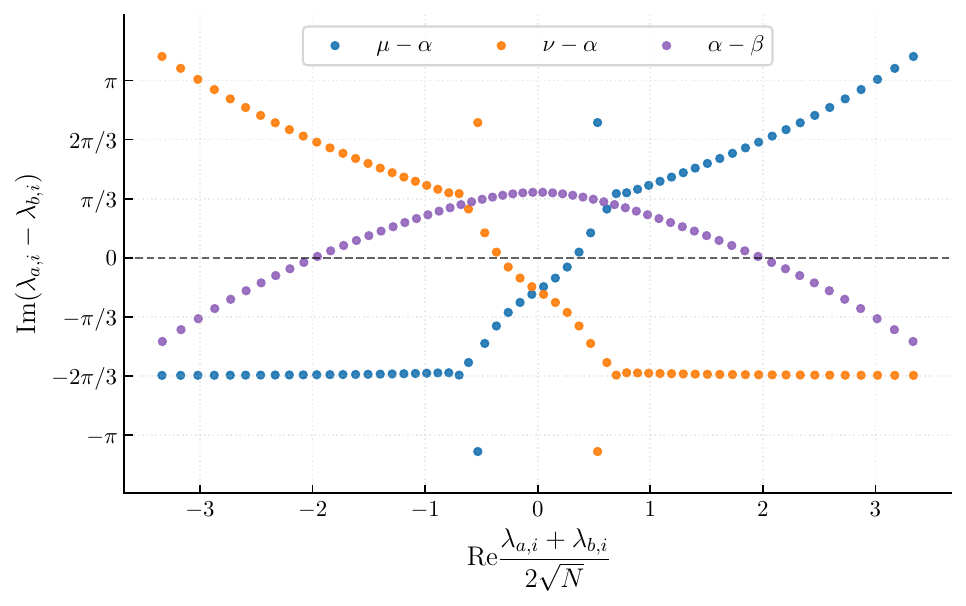}
  \caption{
    Diagnostics of the $Q^{1,1,1}/\bZ_k$ saddle.
    (a) Normalized density $\rho(x)$ of real parts for the zero-level node ($\alpha$) in the rescaled variable $x=\mathrm{Re}\lambda/\sqrt{N}$;
    the $\beta$ node shows an identical profile, and the $\pm k$ nodes ($\mu,\nu$) follow the same envelope within numerical precision.
    (b) Imaginary-part differences along quiver links versus the average real part.
    The link phases show $x$-dependent structure in the bulk with tail plateaus near multiples of $\pi/3$:
    $\mathrm{Im}(\mu-\alpha) \simeq -2\pi/3$ (left tail),
    $\mathrm{Im}(\nu-\alpha) \simeq -2\pi/3$ (right tail),
    while $\mathrm{Im}(\alpha-\beta)$ varies across the support.
    The separations are $\cO(1)$ and $N$ independent.
  }
  \label{fig:Q111_diagnostics}
\end{figure}

Figure~\ref{fig:Q111_diagnostics}(a) shows the normalized density of real parts for the zero-level node,
$
 \rho(x)=\frac{1}{N} \sum_{i=1}^{N}\delta \bigl(x-\frac{\mathrm{Re}\lambda_i}{\sqrt{N}}\bigr) \, ,
$
which forms a smooth, compactly supported profile in the rescaled variable $x=\mathrm{Re}\lambda/\sqrt{N}$.
The distribution is symmetric under $x\!\to\!-x$, so $\rho(x)$ and $\rho(|x|)$ have identical shapes, reflecting the parity symmetry of the saddle about $\mathrm{Re}\lambda=0$.
The conjugate node $\beta$ exhibits the same distribution, while the $\pm k$ nodes $(\mu,\nu)$ follow the same envelope up to small finite-$N$ effects.
Figure~\ref{fig:Q111_diagnostics}(b) presents the imaginary-part differences along the quiver links versus the average real part for the independent combinations $(\mu-\alpha)$, $(\nu-\alpha)$, and $(\alpha-\beta)$.
The remaining two, $(\beta-\nu)$ and $(\beta-\mu)$, coincide numerically with $(\mu-\alpha)$ and $(\nu-\alpha)$, respectively:
for the converged saddles, we find
$\mathrm{Im} ( \beta - \nu ) = \mathrm{Im} ( \mu - \alpha)$ and
$\mathrm{Im} ( \nu - \alpha ) = \mathrm{Im} ( \beta - \mu )$ across all accessible $N$.
The link phases show clear $x$ dependence in the bulk: $\mathrm{Im} ( \mu - \alpha)$ approaches $-2\pi/3$ on the left tail,
$\mathrm{Im} ( \nu - \alpha)$ approaches $ - 2\pi/3$ on the right tail, and $\mathrm{Im} ( \alpha - \beta )$ varies continuously across the support.
These variations remain $\cO (1)$ and stable with $N$, indicating that the imaginary separations are $N$ independent.

\begin{figure}[htp!]
  \centering
  \includegraphics[width=\columnwidth]{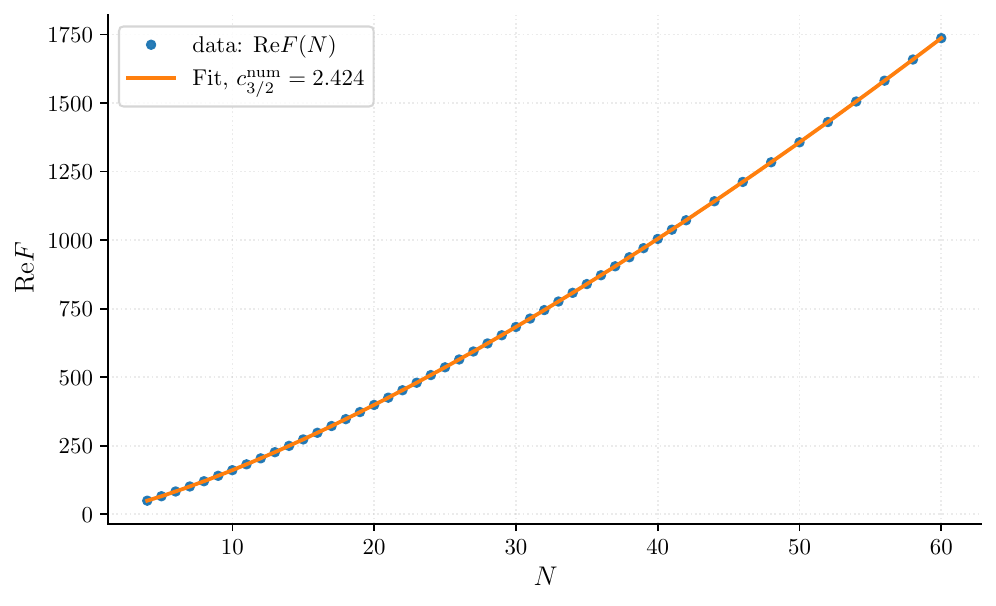}
  \caption{
  Real part of the $S^3$ free energy $F(N)$ for $Q^{1,1,1}/\bZ_k$ (points) with least-squares fit (line).
  The fitted $N^{3/2}$ coefficient agrees with $c_{3/2}^{\text{hol}} = 2.418$ from \eqref{Q111:c32:an}.
  }
  \label{fig:FN_fit}
\end{figure}

The numerical free energy for $Q^{1,1,1}/\bZ_k$ is shown in Fig.~\ref{fig:FN_fit} together with the least-squares fit described above.
The fitted coefficient of the leading $N^{3/2}$ term,
$c_{3/2}^{\text{num}} = 2.424$,
is in excellent agreement with the holographic value
$c_{3/2}^{\text{hol}} = 2.418$ from \eqref{Q111:c32:an}.

We have also explored noncanonical trial R-charge assignments for the $Q^{1,1,1}/\mathbb{Z}_k$ quiver. Away from the canonical point, however, the saddle-point equations become significantly stiffer and the basin of attraction of the physical saddle narrows. With the present implementation, this limits the accessible values of $N$ and prevents a reliable large-$N$ comparison with the holographic prediction in generic noncanonical cases.

We now turn to the $D_3$ quiver theory \eqref{D3:quiver}.
For the canonical R-charges $\Delta_{A_r}=\Delta_{B_r}=\Delta_{C_r}=1/3$,
we obtain convergent solutions up to $N=44$,
and the numerical saddle shows the same qualitative structure as in $Q^{1,1,1}/\bZ_k$:
the eigenvalues form smooth branches in the complex plane,
with real parts that broaden as $\sqrt{N}$ and imaginary parts that remain $\cO(1)$.
The normalized density of real parts is symmetric under $x\!\to\!-x$ and identical across all nodes.
The imaginary components organize into node-dependent bands,
exhibiting approximate plateaus on the tails and mild $x$ dependence in the bulk.
Empirically, we find the pairwise relations
$
\mathrm{Im} ( \mu - \alpha ) = \mathrm{Im} ( \nu - \beta ) \, ,
$
$
\mathrm{Im} ( \alpha - \nu ) = \mathrm{Im} ( \beta - \mu ) \, ,
$
which hold throughout the continuation window.
The fitted coefficient of the leading $N^{3/2}$ term,
$c_{3/2}^{\text{num}}=2.177$,
matches the holographic value
$c_{3/2}^{\text{hol}}=2.280$ within a few percent.

\begin{figure}[htp!]
  \centering
  \includegraphics[width=\columnwidth]{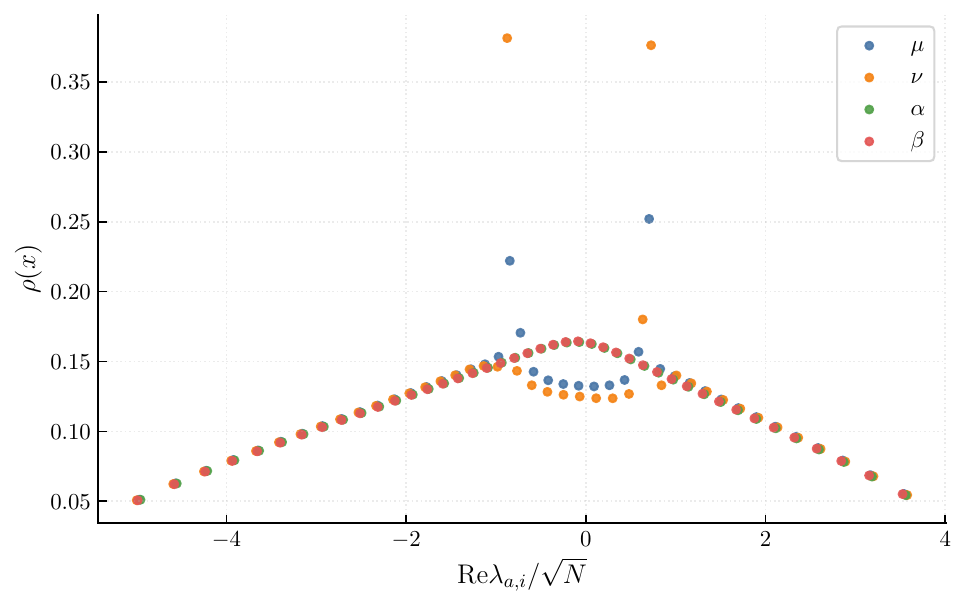}\\[2pt]
  \includegraphics[width=\columnwidth]{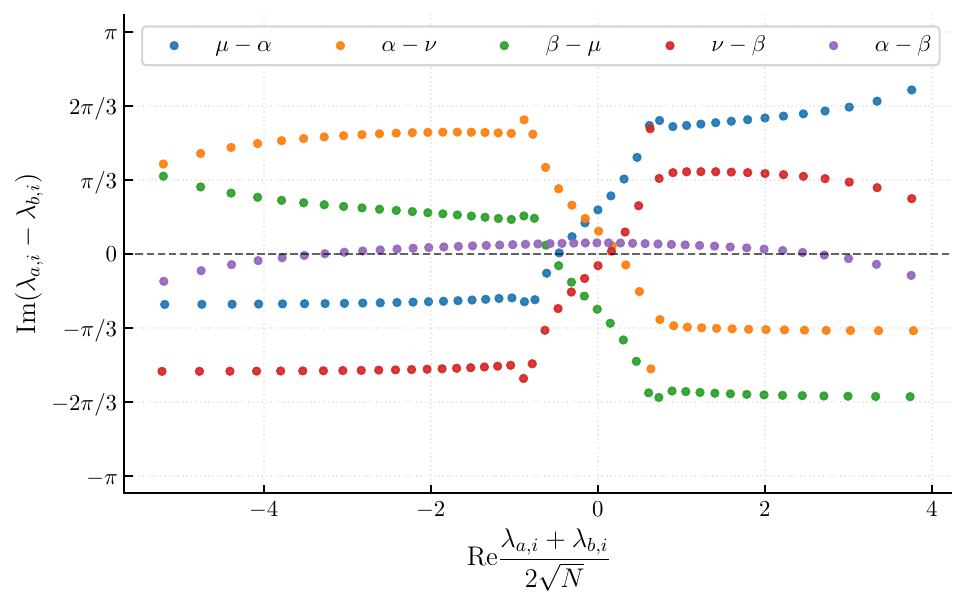}
  \caption{
  Diagnostics for the noncanonical $D_3$ saddle. 
  (a) Normalized density $\rho(x)$ of real parts showing $\sqrt{N}$ broadening in $x=\mathrm{Re}\lambda/\sqrt{N}$. 
  (b) Imaginary-part differences along quiver links versus the average real part. 
  The saddle remains smooth and stable across the continuation range.
  }
  \label{fig:D3_noncanonical}
\end{figure}

For the noncanonical R-charge assignment
$\Delta_{A_1} = 0.12$,
$\Delta_{A_2} = 0.18$,
$\Delta_{C_2} = 0.27$,
$\Delta_{C_1} = 0.33$,
$\Delta_{B_1} = 0.41$, and
$\Delta_{B_2} = 0.69$,
which satisfies the marginality condition but no longer maximizes the free energy,
we obtain stable solutions up to $N=44$.
The holographic benchmark is given by \eqref{D3:c32:an:non-can}.
The numerical saddle remains smooth and stable across the entire continuation range.
As shown in Fig.~\ref{fig:D3_noncanonical},
the real parts follow the same $\sqrt{N}$ broadening as in the canonical case,
while the imaginary components reorganize into distinct node-dependent bands.
The empirical relations observed at the canonical point are no longer present.
The fitted coefficient of the leading $N^{3/2}$ term,
$c_{3/2}^{\text{num}} = 2.048$,
matches the holographic value
$c_{3/2}^{\text{hol}} = 2.084$
within a few percent,
confirming that the large-$N$ scaling and stability of the saddle persist even away from the canonical R-charges.

To further probe the stability of the numerical procedure away from the free energy extremum, we repeated the $D_3$ analysis for ten generic choices of trial R-charges sampled from the region $\Delta_I \in ( 0 , 1 )$ and satisfying the marginality constraint
$\sum_{r=1}^{2}(\Delta_{A_r}+\Delta_{B_r}+\Delta_{C_r})=2$.
In all cases examined, the continuation-and-refinement scheme converges to smooth complex saddles with the same qualitative features as above: the real parts broaden as $\sqrt{N}$, while the imaginary parts remain $\cO(1)$ and arrange into node-dependent bands.
The corresponding large-$N$ fits, performed up to $N = 35$--$50$ depending on the stiffness of each case, yield values of $c_{3/2}$ that agree with the holographic predictions within $5\%$--$10\%$.

As a representative example, for
$\Delta_{A_1}=0.17$, $\Delta_{A_2}=0.10$, $\Delta_{C_2}=0.25$, $\Delta_{C_1}=0.30$, $\Delta_{B_1}=0.28$, and $\Delta_{B_2}=0.90$,
we obtain a fitted large-$N$ coefficient $c_{3/2}^{\text{num}}=1.831$,
in good agreement with the holographic prediction $c_{3/2}^{\text{hol}}=1.788$.
These results indicate that the $N^{3/2}$ scaling and stability of the saddle persist throughout the allowed parameter space of trial R-charges.

\emph{Conclusions and outlook---}%
The main result of this Letter is the first reliable large-$N$ determination of the $S^3$ free energy in chiral $\cN=2$ Chern-Simons-matter theories.
For the $Q^{1,1,1}/\bZ_k$ and $D_3$ quivers, we solve the finite-$N$ saddle-point equations using an adiabatic continuation in $N$: a convergent low-$N$ seed is first obtained with Anderson acceleration, after which $N$ is increased step by step, solving at each stage with a Levenberg-Marquardt or Powell-hybrid update and refining the result with a damped Newton iteration with backtracking. This continuation-and-refinement scheme yields smooth complex saddles whose real parts scale as $\sqrt{N}$ while their imaginary parts remain $\cO(1)$.
The resulting $F \sim N^{3/2}$ agrees quantitatively with the holographic prediction~\eqref{AdS:CFT}, resolving the long-standing mismatch in the large-$N$ behavior of chiral quivers and restoring consistency with AdS$_4$/CFT$_3$.

The eigenvalue profiles suggest several directions for further work.
First, the observed saddle structure---$\sqrt{N}$ support, $\cO(1)$ imaginary separations, and the smooth density and link-phase functions---provides the essential input for an \emph{analytic} solution of the continuum saddle-point equations, and hence a direct derivation of the large-$N$ free energy for chiral gauge theories.
Second, it is natural to revisit the proposed relation between eigenvalue distributions and operator counting in the chiral ring \cite{Gulotta:2011si,Gulotta:2011aa}.
For $Q^{1,1,1}/\bZ_k$, both our numerical saddle and the operator-counting analysis of \cite{Kim:2012vza} yield a triangular real-part density, but the link-phase profiles differ, indicating that the relation between eigenvalues and operator counting observed in nonchiral theories does not extend straightforwardly to the chiral case.
We will return to this comparison in future work.
Third, distinct nonchiral quivers for $\cC(Q^{1,1,1}/\bZ_k)$ and for $D_3$ were proposed in \cite{Benini:2009qs}.
Since our large-$N$ free energies match those of the nonchiral theories in \cite{Jafferis:2011zi}, this provides the first quantitative evidence that the chiral and nonchiral descriptions flow to the same infrared fixed point.

Our scheme need not access \emph{all} saddles.
For other chiral quivers dual to AdS$_4\times M^{1,1,1}$ and AdS$_4\times Q^{2,2,2}$, we did not find $N^{3/2}$ scaling; this may indicate additional complex saddles outside the basin of attraction of our procedure, or the absence of an M-theory phase for those Chern-Simons quivers at large $N$.

More broadly, the same framework can be used to compute supersymmetric indices that capture the microstate degeneracy of AdS$_4\times Y_7$ black holes.

\medskip

\emph{Acknowledgments---}%
I thank Amihay Hanany and Costis Papageorgakis for valuable discussions.
I am especially grateful to Alberto Zaffaroni for insightful comments on a draft of this work and for many illuminating conversations.
I also thank my wife, Parastoo Salah, for convincing me to switch from \texttt{Mathematica} to \texttt{Python}, and for her patience with my repeated ``just one more question'' questions.
I thank the referees for their careful reading of the manuscript and for their constructive comments, which helped improve the clarity and scope of this work.
This research was supported by UK Research and Innovation (UKRI) under the UK government's Horizon Europe funding guarantee (Grant. No. EP/Y027604/1).

\bibliography{3dchiral}

\end{document}